\documentclass[11pt]{article}
\usepackage{axodraw}
\usepackage{epsfig}
\usepackage{amsfonts}
\usepackage{amsmath}
\usepackage{bbm}
\allowdisplaybreaks[1]

\input pix.sty

\newcommand{\mW}{m_\rmii{$W$}}
\newcommand{\mZ}{m_\rmii{$Z$}}
\newcommand{\mH}{m_\rmii{$H$}}
\def\undertilde#1{\mathop{\vtop{\ialign{##\cr$\textstyle{#1}$\cr%
\noalign{\kern1pt\nointerlineskip}\hfil$\mathchar"0365$\hfil\cr}}}}
\def\wideundertilde#1{\mathop{\vtop{\ialign{##\cr$\textstyle{#1}$\cr%
\noalign{\kern1pt\nointerlineskip}\hfil$\mathchar"0367$\hfil\cr}}}}

\renewcommand{\eq}{eq.~}
\renewcommand{\eqs}{eqs.~}
\renewcommand{\se}{sec.~}

\renewcommand{\fig}{fig.~}

\newcommand{\Nc}{N_{\rm c}}

\newcommand{\gammaE}{\gamma_\rmii{E}}

\newcommand{\rmO}{{\mathcal{O}}}
\newcommand{\bmu}{\bar\mu}

\def\lsi{\raise0.3ex\hbox{$<$\kern-0.75em\raise-1.1ex\hbox{$\sim$}}}
\def\gsi{\raise0.3ex\hbox{$>$\kern-0.75em\raise-1.1ex\hbox{$\sim$}}}
\newcommand{\lsim}{\mathop{\lsi}}
\newcommand{\gsim}{\mathop{\gsi}}

\newcommand{\nF}{n_\rmii{G}} % {n_\rmii{F}}

\newcommand{\rmii}[1]{{\mbox{\tiny\rm{#1}}}}

\newcommand{\Tint}[1]{{\hbox{$\sum$}\!\!\!\!\!\!\!\int\,}_{\!\!\!\!\raise-0.9ex\hbox{$\scriptstyle{#1}$}}}
\newcommand{\Tinti}[1]{{{\Sigma}\!\!\!\!\raise0.3ex\hbox{$\int$}_\rmii{${#1}$}}}

\newcommand{\bi}{\begin{itemize}}
\newcommand{\ei}{\end{itemize}}
\newcommand{\hide}[1]{ }

\def\TAsc(#1,#2)(#3,#4,#5)%
{\SetWidth{2.0}\CArc(#1,#2)(#3,#4,#5)\SetWidth{1.0}}
\def\Lwidth{3}

\def\TAgl(#1,#2)(#3,#4,#5){\SetWidth{2.0}\PhotonArc(#1,#2)(#3,#4,#5){\Lwidth}%
{6.283 #3 mul 360 div #4 #5 sub #4 #5 sub mul sqrt mul Tdensity mul}%
\SetWidth{1.0}}
\def\TLgl(#1,#2)(#3,#4){\SetWidth{2.0}\Photon(#1,#2)(#3,#4){\Lwidth}
{#1 #3 sub #1 #3 sub mul #2 #4 sub #2 #4 sub mul add sqrt Tdensity mul}%
\SetWidth{1.0}}
\renewcommand{\picc}[1]{\;\parbox[c]{60pt}{\begin{picture}(60,30)(0,0)
\SetWidth{1.0}\SetScale{1.3} #1 \end{picture}}\; }
\def\Lwidth{1.3}

\newcommand{\picu}[1]{\;\parbox[c]{40pt}{\begin{picture}(50,30)(0,0)
\SetWidth{1.0}\SetScale{0.8} #1 \end{picture}}\; }

\def\ScatA{\picu{%
 \SetWidth{2.0} 
 \DashCArc(30,20)(30,0,360){3}%
 \Asc(30,-8)(41,42,138)%
 \Asc(30,48)(41,222,318)%
 \GCirc(30,-10){3}{0}%
}}
\def\ScatB{\picu{%
 \SetWidth{2.0} 
 \DashCArc(30,20)(30,0,360){3}%
 \Asc(30,20)(10,0,360)%
 \GCirc(30,-10){3}{0}%
 \Photon(0,20)(20,20){2}{3}%
 \Photon(40,20)(60,20){2}{3}%
}}
\makeatletter \@addtoreset{equation}{section} \makeatother
\renewcommand{\theequation}{\arabic{section}.\arabic{equation}}
\makeatletter
\renewcommand\section{\@startsection {section}{1}{\z@}%
                                   {-5.5ex \@plus -1ex \@minus -.2ex}% bfr-
                                   {2.3ex \@plus.2ex}%
                                   {\normalfont\large\bfseries}}
\renewcommand\subsection{\@startsection{subsection}{2}{\z@}%
                                     {-3.25ex\@plus -1ex \@minus -.2ex}%
                                     {1.5ex \@plus .2ex}%
                                     {\normalfont\normalsize\bfseries}}
\renewcommand\thesection {\@arabic\c@section}
\renewcommand\thesubsection   {\thesection.\@arabic\c@subsection}
\renewcommand{\@seccntformat}[1]{%
\csname the#1\endcsname.\hspace{1.0em}}
\makeatother
\begin{document}

\flushbottom

\begin{titlepage}
\begin{flushright}
% OUTLINE / M.L. \\
%arXiv:yymm.nnnn\\ 
\vspace*{1cm}
\end{flushright}
\begin{centering}
\vfill

{\Large{\bf
 Standard Model thermodynamics across \\[3mm] the electroweak crossover
}} 

\vspace{0.8cm}

M.~Laine %$^{\rm a}$ %%\footnote{laine@itp.unibe.ch}
and 
M.~Meyer %$^{\rm a}$ %%\footnote{meyer@itp.unibe.ch}

\vspace{0.8cm}

%$^\rmi{a}$%

{\em
Institute for Theoretical Physics, 
Albert Einstein Center, University of Bern, \\
  %for Fundamental Physics, 
Sidlerstrasse 5, CH-3012 Bern, Switzerland
}
 
\vspace*{0.8cm}

\mbox{\bf Abstract}
 
\end{centering}

\vspace*{0.3cm}
 
\noindent
Even though the Standard Model with a Higgs mass $\mH = 125$~GeV possesses
no bulk phase transition, its thermodynamics still experiences a ``soft 
point'' at temperatures around $T = 160$~GeV, with a deviation from ideal gas 
thermodynamics. Such a deviation may have an effect on precision computations 
of weakly interacting dark matter relic abundances if their mass is in 
the few TeV range, or on leptogenesis scenarios operating in this temperature
range. By making use of results from lattice simulations based on a 
dimensionally reduced effective field theory, we estimate the relevant 
thermodynamic functions across the crossover. The results are tabulated in 
a numerical form permitting for their insertion as a background equation of 
state into cosmological particle production/decoupling codes. We find that
Higgs dynamics induces a non-trivial ``structure'' visible e.g.\ in the
heat capacity, but that in general the largest radiative corrections 
originate from QCD effects, reducing the energy density by a couple of 
percent from the free value even at $T > 160$~GeV. 

\vfill

%\noindent
%PACS numbers: 
%11.10.Wx, %        Finite temperature field theory
%11.15.Ha, %        Lattice gauge theory 
%98.80.Cq, % 	    Particle-theory models of the early Universe 
%\\
%Keywords:
 
\vspace*{1cm}
  
\noindent
June 2015

\vfill

\end{titlepage}

%%%%%%%%%%%%%%%%%%%%%%%%%%% SECTION %%%%%%%%%%%%%%%%%%%%%%%%%%%%%%%%%%%%%%
%
\section{Introduction}
\la{se:intro}

The excellent performance of the Standard Model (SM) in describing LHC data
suggests that the SM represents a precise description of nature
up to energy scales of several hundred GeV. If this observation 
continues to be confirmed by future runs, one consequence 
is that the Higgs phenomenon set in smoothly, i.e.\ 
without a phase transition, as the Universe cooled down to temperatures 
below 160~GeV~\cite{endpoint}--\cite{ep4}
(cf.\ ref.~\cite{review} for a review of refined investigations). 
This would mean that the bulk motion of the plasma did not
depart from thermal equilibrium and therefore did not generate 
cosmological relics. 

Nevertheless, the thermodynamics of the SM
could still have left a ``background'' imprint on other non-equilibrium
physics that just happened to be going on. 
One example is that the B+L violating
rate switched off rapidly around the crossover~\cite{anders1}, 
and therefore determined which fraction of a given lepton number 
being produced around these temperatures
could be converted into baryons 
(cf.\ e.g.\ ref.~\cite{numsm_rev}). 
Another is that if Dark Matter particles
decoupled at around this period, even small features 
in the equation of state could have had an impact~\cite{steigman}. 
(Similar considerations can be carried out for the QCD crossover, 
cf.\ refs.~\cite{qcd1,qcd2} and references therein.)
A rough estimate for the decoupling temperature of weakly
interacting particles of mass $M$ is $T \sim M/25$, so that 
we may expect the largest sensitivity for $M\sim 4$~TeV or so.  

The purpose of our study is to estimate the equation of state
of the SM around the electroweak crossover. We do this through 
perturbative computations extending up to 3-loop
level, as well as by making use of existing lattice simulations within 
a dimensionally reduced effective theory. Only the temperature is 
treated as a non-trivial canonical variable, 
with chemical potentials associated with 
conserved charges such as the baryon minus lepton number or 
the hypercharge magnetic flux set to zero. 

The paper is organized as follows. 
After defining the basic observables in~\se\ref{se:setup}, 
we derive a ``master equation'' in~\se\ref{se:eos} which relates the
trace of the expectation value of the energy-momentum tensor to 
three ingredients. The first ingredient, scale violations
through quantum corrections, is addressed in \se\ref{se:quantum}.
The second ingredient, the expectation value of the Higgs condensate, 
needs to be evaluated up to the non-perturbative level, and this 
is achieved in \se\ref{se:pdp}. The third ingredient is related 
to vacuum renormalization and is discussed in~\se\ref{se:vac}. 
All ingredients are put together in~\se\ref{se:all}, where we 
also present phenomenological results. 
Section~\ref{se:concl} contains some conclusions as well as a discussion
of the theoretical
uncertainties of the current analysis. Readers not interested in technical
details may start from \se\ref{se:all}.

%%%%%%%%%%%%%%%%%%%%%%%%%%% SECTION %%%%%%%%%%%%%%%%%%%%%%%%%%%%%%%%%%%%%%
%
\section{Basic setup}
\la{se:setup}

We consider the Standard Model with a Higgs potential of the form
\be
 \delta \mathcal{L}^{ }_\rmii{E} = 
 -\nu^2_\rmii{B} \phi^\dagger \phi + 
 \lambda^{ }_\rmii{B} (\phi^\dagger\phi)^2
 \;, 
\ee
where $\nu^2_\rmii{B}$, $\lambda^{ }_\rmii{B}$ are bare parameters
(the corresponding renormalized parameters are denoted by 
$\nu^2$, $\lambda$), and  $\mathcal{L}^{ }_\rmii{E}$ is 
a Euclidean (imaginary-time) Lagrangian. 
Denoting by 
$
 g^2 \!\in\! \{ \lambda, h_t^2, g_1^2, g_2^2, g_3^2 \}
$ 
a generic coupling constant\footnote{% 
 Here $h_t$ denotes the top Yukawa coupling 
 and $g_1,g_2,g_3$ are the gauge couplings related
 to the Standard Model gauge groups 
 U$^{ }_\rmii{Y}(1)$, SU$^{ }_\rmii{L}(2)$ and 
 SU$^{ }_\rmii{C}(3)$.
 }, 
we concentrate on the parametric temperature range 
\be
 T^2 \; \gsim \; \frac{\nu^2}{g^2}
 \la{nu2}
\ee
in the following. At the lower edge of this range, 
the effective Higgs mass parameter of the 
dimensionally reduced theory~\cite{dr1,dr2}, $\bar{m}_3^2$~\cite{generic}, 
is assumed to satisfy
\be
 | \bar{m}_3^2 | 
 \sim
 \bigl|  -\nu^2 + g^2 T^2 \bigr|
 \; \lsim \; \frac{g^3 T^2}{\pi}
 %  \sim \Bigl ( \frac{g^2 T}{\pi} \Bigr ) ^2
 \;, \la{range}
\ee
but it can have either sign. If it is negative, we may 
expect to find ourselves in a ``Higgs phase''. Within
a gauge-fixed perturbative treatment, this would imply
that the Higgs field has an expectation value  
$v^2 \sim - \bar{m}_3^2/\lambda > 0$. Within the range 
of \eq\nr{range}, only relatively small expectation values
$v \lsim g^{\fr12}T$ can be considered, where
we counted $\lambda \sim g^2$.

It is well known that when momenta in the range 
$
  | \bar{m}_3^2 | \sim (g^2 T/\pi)^2
$
are considered, which is a special case of 
\eq\nr{range}, then the dynamics of the 
system is non-perturbative~\cite{linde,gpy}. Therefore
the dynamics needs to be treated with lattice methods. However, 
non-perturbativity is associated with particular 
modes only, and can be captured with a dimensionally 
reduced effective description~\cite{dr1,dr2}. 
Even though the construction of this theory is perturbative, we expect 
to obtain a description accurate on the percent level within a weakly coupled
theory such as the SM. This estimate is based on analyzing the influence from
higher-order operators that are truncated from the dimensionally reduced 
theory (cf.\ sec.~5.4 of ref.~\cite{generic}) 
and, more concretely, from a comparison
of non-perturbative results for the location of the endpoint in the phase
diagram of an SM-like theory based on the dimensionally reduced
description and on direct 4-dimensional lattice simulations~\cite{4d,gR2}. 
(It would be interesting to repeat the latter type of a comparison for the 
physical value of the Higgs mass.)

The basic observable that we consider is the thermodynamic pressure
of the SM. The pressure can formally be defined through 
the grand canonical partition function $\mathcal{Z}$ as
\be
 \mathcal{Z} \equiv 
 \exp \Bigl( {\frac{p^{ }_\rmii{B}(T) V}{T}} \Bigr) 
 \;, \la{p_def}
\ee
where the thermodynamic limit $V \to \infty$ is implied, 
and $p^{ }_\rmii{B}$ denotes the bare result.  
The computation of $p^{ }_\rmii{B}$ has previously been considered in 
ref.~\cite{gv1} for the case $\bar{m}_3^2 \sim g^2 T^2$, 
and in ref.~\cite{gv2} for $\bar{m}_3^2 \sim g^3 T^2 /\pi $.
Following 
refs.~\cite{gv1,gv2}\footnote{%
 We agree with most of the results in these papers, however disagree
 with the renormalization condition for $p$
 and with certain technical details, 
 cf.\ appendix~B. 
 } (which were inspired by ref.~\cite{bn})
we write
\be
 p^{ }_\rmii{B}(T) = p^{ }_\rmii{E}(T) + p^{ }_\rmii{M}(T)
      + p^{ }_\rmii{G}(T)
 \;, \la{splitup}
\ee
where $p^{ }_\rmii{E}$ collects contributions from the 
momentum scale $k \sim \pi T$, $p^{ }_\rmii{M}$  those from 
$k \sim gT$, $p^{ }_\rmii{G}$ those from $k \sim g^2 T/\pi$. 
Rephrasing terminology often used in the QCD context, we refer
to the effective theory contributing
to $p^{ }_\rmii{M}$ as ESM (``Electrostatic Standard Model'') and to that 
contributing to $p^{ }_\rmii{G}$ as MSM (``Magnetostatic Standard Model'').

As defined
by \eq\nr{p_def} the pressure is ultraviolet divergent. 
We renormalize it by assuming that the pressure (and energy density)
vanish at $T = 0$. 
The renormalized pressure can then be written as 
\be
 p^{ }_{ }(T) \equiv p^{ }_\rmii{B}(T) - p^{ }_\rmii{B}(0) 
 \;. \la{renorm}
\ee
Our goal is to determine the dimensionless functions 
$p^{ }_{ }(T) / T^4$ and 
$T \partial_T \{\, p^{ }_{ }(T) / T^4\, \}$ 
up to $\rmO(g^5)$. For the latter quantity, 
it turns out that the contribution of the softest momentum
modes needs to be 
treated non-perturbatively in order to reach this precision. 

%%%%%%%%%%%%%%%%%%%%%%%%%%% SECTION %%%%%%%%%%%%%%%%%%%%%%%%%%%%%%%%%%%%%%
%
\section{Master equation}
\la{se:eos}

Consider a normalized form of \eq\nr{renorm}, 
$
 p/T^4 = [p^{ }_\rmii{B}(T) - p^{ }_\rmii{0B}]/T^4
$, 
where $p^{ }_\rmii{0B} \equiv p^{ }_\rmii{B}(0)$ 
denotes the bare zero-temperature pressure. 
Unless stated  otherwise ultraviolet divergences are treated through
dimensional regularization. In the difference of \eq\nr{renorm}
all $1/\epsilon$ poles cancel, so that the bare expressions
can be replaced with renormalized expressions in the
$\msbar$ scheme ($p^{ }_\rmii{B}\to p^{ }_\rmii{R}$, 
$p^{ }_\rmii{0B}\to p^{ }_\rmii{0R}$): 
\ba
 \frac{p^{ }_{ }(T)}{T^4} & = &  
 \frac{ p^{ }_\rmii{R} (T; \bmu, \nu^2(\bmu),g^2(\bmu)) - 
 p^{ }_\rmii{0R}(\bmu, \nu^2(\bmu),g^2(\bmu))}{T^4}
 \;. \la{renorm2}
\ea
Here $\bmu$ denotes the $\msbar$ renormalization scale.

We envisage that at some temperature
($T = T^{ }_0$) on either side of the crossover, 
say $T^{ }_0 \ll 160$~GeV or 
$T^{ }_0 \gg 160$~GeV so that 
$ 
 | \bar{m}_3^2 | \gsim {g^3 T^2} / {\pi}
$, 
$p^{ }_{ }/T^4$ can be determined by a direct perturbative computation. 
The task then
is to integrate $p^{ }_{ }/T^4$ across the electroweak crossover:
\be
  \frac{p^{ }_{ }(T^{ }_1)}{T_1^4} - 
  \frac{p^{ }_{ }(T^{ }_0)}{T_0^4}
  = 
 \int_{T^{ }_0}^{T^{ }_1} \! \frac{{\rm d}T}{T}
 \; \times
 \; T \frac{{\rm d}}{{\rm d}T} 
 \biggl\{ \frac{p^{ }_{ }(T)}{T^4} \biggr\}
 \;. \la{integral_method}
\ee
So, we need to compute the logarithmic temperature derivative
of the dimensionless ratio $p^{ }_{ } / T^4$. The result is non-zero
because of the breaking of scale invariance, either by the explicit
mass term $\nu^2$, or by quantum corrections related to running couplings. 

Making use of standard thermodynamic relations, we note that
\be
 T \frac{{\rm d}}{{\rm d}T} 
 \biggl\{ \frac{p^{ }_{ }(T)}{T^4} \biggr\} 
 = \frac{e^{ }_{ }(T) - 3 p^{ }_{ }(T)}{T^4}
 \; \equiv \; \Delta(T)
 \;, 
\ee
where $e$ denotes the energy density. 
In the context of QCD this quantity is referred to as the trace anomaly, 
and has been studied with lattice methods. Were it not 
that chiral fermions (in particular the top quark) are essential for
the physics considered, the
problem could in principle be studied with full 4-dimensional 
lattice simulations here as well. 
In the present paper, we circumvent the problem
of chiral fermions by using lattice input only for the 
Bose-enhanced infrared degrees of freedom, treating fermions within
a (resummed) weak-coupling expansion.  

Now, because of dimensional reasons, we may rewrite \eq\nr{renorm2} as
\ba
 \frac{p^{ }_{ }(T)}{T^4} & \equiv & 
 \hat{p}^{ }_\rmii{R}
 \Bigl( \frac{\bmu}{T}, \frac{\nu^2(\bmu)}{T^2}, g^2(\bmu) \Bigr)
 - \frac{p^{ }_\rmii{0R}}{T^4} \bigl( \bmu , \nu^2(\bmu),g^2(\bmu) \bigr)
 \;.
\ea
We note from
the Euclidean path integral representation, {\em viz.} 
\be
 \mathcal{Z} = \int \! \mathcal{D} [...] \,
 \exp
 \Bigl\{- \int_V \! {\rm d}^{3-2\epsilon}\vec{x} \int_0^{1/T} \! {\rm d}\tau
 \mathcal{L}^{ }_\rmii{E}
 \Bigr\}
 \;, 
\ee
that the bare pressure obeys 
(here $\hat{p}^{ }_\rmii{B} \equiv p^{ }_\rmii{B}/T^4$)
\be
 \frac{\partial \hat{p}^{ }_\rmii{B} }{\partial(\nu^2(\bmu)/T^2)}
 = \frac{
 [ \mathcal{Z}_m \langle \phi^\dagger \phi \rangle ]^{ }_\rmii{B} }{T^2}
 \;, \la{pdp}
\ee
where we wrote
$
 \nu^2_\rmii{B} = \mathcal{Z}_m\, \nu^2(\bmu)
$.\footnote{%
 The renormalization factor reads
 $
  \mathcal{Z}_m = 
  1 + \frac{3}{(4\pi)^2\epsilon}
  \bigl[
    2 \lambda + h_t^2 - \fr14 (g_1^2 + 3 g_2^2) 
  \bigr] + \rmO(g^4)
 $
 but is not separately needed, because it always appears
 in the combination
 $
  \mathcal{Z}_m \langle \phi^\dagger \phi \rangle
 $.
 } 
It is now straightforward to obtain the following relation:
\be
 T \frac{{\rm d}}{{\rm d}T}
 \biggl\{ \frac{p^{ }_{ }(T)}{T^4} \biggr\}
 = -
 \frac{\partial \hat{p}^{ }_\rmii{R} }{\partial \ln(\bmu/T)}
 - \frac{2 \nu^2(\bmu) \, 
 [ \mathcal{Z}_m \langle \phi^\dagger \phi \rangle ]^{ }_\rmii{R} }{T^4}
 + \frac{4 p^{ }_\rmii{0R}}{T^4}
 \;. \la{master}
\ee
Here eq.~\nr{pdp} has been replaced by its renormalized version.

It can be observed from \eq\nr{master} that three ingredients
are needed: the determination of 
explicit logarithms appearing in $\hat{p}^{ }_\rmii{R}$
(``breaking of scale invariance by quantum corrections''); 
the temperature evolution
of the Higgs condensate
(``explicit breaking of scale invariance''); 
as well as the vacuum term needed for 
renormalization. We discuss the first of these in the next section, 
the condensate in \se\ref{se:pdp}, and the vacuum term
in \se\ref{se:vac}. The results are collected together
and evaluated numerically in \se\ref{se:all}.

%%%%%%%%%%%%%%%%%%%%%%%%%%% SECTION %%%%%%%%%%%%%%%%%%%%%%%%%%%%%%%%%
%
\section{Effects from scale violation by quantum corrections}
\la{se:quantum}

The first ingredient needed in \eq\nr{master} are logarithms of 
$\bmu/T$ that are induced by loop corrections. 
There are two ways to extract 
$
 \frac{\partial \hat{p}^{ }_\rmii{R}}{\partial \ln(\bmu/T)} 
$
from ref.~\cite{gv1}:
either by reading logarithms from the explicit expressions for the 
various coefficients given there, or by deducing them from the running
of the couplings. Making use of the notation in \eq\nr{splitup},
with $\hat{p} \equiv p/T^4$, the 
part $\hat{p}^{ }_\rmii{G}$ contributes at $\rmO(g^6)$.
{}From $\hat{p}^{ }_\rmii{E} + \hat{p}^{ }_\rmii{M}$ 
we need terms up to and including $\rmO(g^5)$.
The expansion reads
\ba
 \hat{p}^{ }_\rmii{E} + \hat{p}^{ }_\rmii{M}
 \!\! & = & \!\!
 \alpha^{ }_\rmii{E1}
 + g_1^2 \alpha^{ }_\rmii{EB}
 + g_2^2 \alpha^{ }_\rmii{EA}
 + g_3^2 \alpha^{ }_\rmii{EC}
 + \lambda \alpha^{ }_\rmii{E$\lambda$}
 + h_t^2 \alpha^{ }_\rmii{EY}
 + \frac{\nu^2}{T^2} \alpha^{ }_\rmii{E$\nu$}
 \la{pEpM} \\ 
 \!\! & + & \!\! 
 \frac{\nu^4}{(4\pi)^2 T^4}
 \biggl(\frac{1}{\epsilon} + \alpha^{ }_\rmii{E$\nu\nu$} \biggr)
 +
 \frac{1}{12\pi T^3} \Bigl[ 
   \bigl( m_\rmii{E1}^{2\,(0)} \bigr)^{\fr32}
   + 3 \bigl( m_\rmii{E2}^{2\,(0)} \bigr)^{\fr32}
   + 8 \bigl( m_\rmii{E3}^{2\,(0)} \bigr)^{\fr32}
 \Bigr]
 + \ldots \;, \hspace*{6mm} \nonumber
\ea
where the  leading-order  Debye masses read 
($\nF = 3$ denotes the number of generations)
\be
 m^{2\,(0)}_\rmii{E1} \; = \; \Bigl( \fr16 + \frac{5\nF}{9} \Bigr) g_1^2 T^2 
 \;, \quad
 m^{2\,(0)}_\rmii{E2} \; = \; \Bigl( \fr56 + \frac{\nF}{3} \Bigr) g_2^2 T^2
 \;, \quad
 m^{2\,(0)}_\rmii{E3} \; = \; \Bigl( 1 + \frac{\nF}{3} \Bigr) g_3^2 T^2
 \;.  \la{mEmasses2}
\ee
All coefficients in \eq\nr{pEpM} apart from $\alpha^{ }_\rmii{E$\nu\nu$}$
are scale independent; its value differs from that in ref.~\cite{gv1} 
because of our different renormalization condition: 
\be
 \alpha^{ }_\rmii{E$\nu\nu$} = d^{ }_\rmii{F}  n^{ }_\rmii{S}
 \biggl(
  % \frac{1}{2\epsilon} + 
  \ln \frac{\bmu}{4\pi T} + \gammaE 
 \biggr)
 \;, \la{aEnn}
\ee
where $d^{ }_\rmii{F},  n^{ }_\rmii{S}$ are specified in \eq\nr{group}.
Given that the pressure as a whole is scale independent, running
from the couplings must cancel against explicit logarithms; therefore, 
\ba
 && \hspace*{-1.5cm}
 \frac{\partial 
 [ \hat{p}^{ }_\rmii{E} + \hat{p}^{ }_\rmii{M}]^{ }_\rmii{R} }
 {\partial \ln(\bmu/T)}
  \; =  \;
 - \, 
   \alpha^{ }_\rmii{EB}\, \bmu \frac{{\rm d} g_1^2}{{\rm d}\bmu} 
 - 
   \alpha^{ }_\rmii{EA}\, \bmu \frac{{\rm d} g_2^2}{{\rm d}\bmu}  
 - 
   \alpha^{ }_\rmii{EC}\, \bmu \frac{{\rm d} g_3^2}{{\rm d}\bmu}  
 -  \alpha^{ }_\rmii{E$\lambda$}\, \bmu \frac{{\rm d} \lambda}{{\rm d}\bmu}
 - \alpha^{ }_\rmii{EY}\, \bmu \frac{{\rm d} h_t^2}{{\rm d}\bmu} 
 -  \frac{\alpha^{ }_\rmii{E$\nu$} }{T^2}\,
    \bmu \frac{{\rm d} \nu^2}{{\rm d}\bmu}
 \nn 
 & + & \frac{\nu^4 }{(4\pi)^2 T^4 } 
    \, \bmu \frac{{\rm d} \alpha^{ }_\rmii{E$\nu\nu$} }{{\rm d}\bmu}
 - \sum_{i=1}^3 
  \frac{ d^{ }_i m_\rmii{E$i$}^{(0)} }{8\pi T^3} 
    \, \bmu
  \frac{{\rm d}
     m_\rmii{E$i$}^{2\,(0)} }{{\rm d}\bmu}
 + \rmO(g^6) \;, \la{dpEpM}
\ea
where $d_1 \equiv 1$, $d_2 \equiv 3$, and $d_3 \equiv 8$.
The runnings can be read from \eqs(7)--(12) of ref.~\cite{gv1}.
Putting together, we obtain
\ba
 - \frac{  \partial
 [ \hat{p}^{ }_\rmii{E} +  \hat{p}^{ }_\rmii{M} ]^{ }_\rmii{R} }
 {\partial \ln(\bmu/T)} & = & \Delta^{ }_1(T) \;, \\
 \Delta^{ }_1(T) & \equiv &  \frac{1}{(4\pi)^2} \biggl\{  
 \frac{198+141\nF-20\nF^2}{54} g_3^4 + 
 \frac{266+163\nF-40\nF^2}{288} g_2^4
 \nn & & \; -  \, 
 \frac{144+375\nF+1000\nF^2}{7776} g_1^4 - \frac{g_2^2 g_1^2}{32} 
 -
 h_t^2 \Bigl( \frac{7h_t^2}{32} - \fr{5g_3^2}6 - \frac{15g_2^2}{64}  - 
 \frac{85g_1^2}{576}   \Bigr) 
 \nn & & \; - \, %% -
 \lambda\Bigl( \lambda + \fr{h_t^2}2 - \frac{g_1^2+3 g_2^2}{8} \Bigr)
 +  %% \nn & & \; + \, 
 \frac{\nu^2}{T^2} 
 \Bigl( h_t^2 + 2 \lambda -  \frac{g_1^2+3 g_2^2}{4} \Bigr)
 - \frac{2 \nu^4}{T^4} 
 \biggr\} 
 \nn 
 & - & 
 \frac{1}{(4\pi)^3}
 \biggl\{
  32 g_3^5
  \Bigl( 1 + \frac{\nF}{3} \Bigr)^{\fr32}
  \Bigl(\frac{11}{4} - \frac{\nF}{3} \Bigr) 
  + 12 g_2^5
  \Bigl( \fr56 + \frac{\nF}{3} \Bigr)^{\fr32}
  \Bigl(\frac{43}{24} - \frac{\nF}{3} \Bigr) 
 \nn & & \; - \, %% -
   4 g_1^5
  \Bigl( \fr16 + \frac{5\nF}{9} \Bigr)^{\fr32}
  \Bigl(\frac{1}{24} + \frac{5\nF}{9} \Bigr) 
 \biggr\}
 + \rmO(g^6)\;. \la{t12}
\ea
This expression is renormalization group (RG) invariant up to $\rmO(g^6)$. 
The numerically largest corrections originate from terms involving the 
strong gauge coupling $g_3^2$.

%%%%%%%%%%%%%%%%%%%%%%%%%%% SECTION %%%%%%%%%%%%%%%%%%%%%%%%%%%%%%%%%%%%%%
%
\section{Effects from the Higgs condensate}
\la{se:pdp}

%%%%%%%%%%%%%%%%%%%%%%%%%%% SUBSECTION %%%%%%%%%%%%%%%%%%%%%%%%%%%%%%%%%%%
%
\subsection{Outline}

The Higgs condensate can be obtained from \eq\nr{pdp}. 
For a non-perturbative evaluation, we make use of the lattice 
simulations in refs.~\cite{anders0,anders1}. This means that we split 
the pressure into contributions from various momentum scales like in 
\eq\nr{splitup}: 
\be
 \mathcal{Z}_m \langle \phi^\dagger\phi\rangle = 
 \frac{\partial p^{ }_\rmii{E}}{\partial \nu^2}
 + \frac{\partial p^{ }_\rmii{M}}{\partial \nu^2}
 + \frac{\partial p^{ }_\rmii{G}}{\partial \nu^2}
 \;. \la{pdp_split}
\ee
Given that $\mathcal{Z}_m \langle \phi^\dagger\phi\rangle$ is 
multiplied by $\nu^2$ in \eq\nr{master} and that 
we assume $\nu^2\sim g^2 T^2$, we only need to determine
$\mathcal{Z}_m \langle \phi^\dagger\phi\rangle$ 
to $\rmO(g^3)$. That said, it turns out that computations
going beyond those in refs.~\cite{gv1,gv2} are needed. 

%%%%%%%%%%%%%%%%%%%%%%%%%%% SUBSECTION %%%%%%%%%%%%%%%%%%%%%%%%%%%%%%%%%%%
%
\subsection{Perturbative contributions}

The first term of \eq\nr{pdp_split}, 
the contribution from the ``hard'' scales $k\sim \pi T$, 
can be directly extracted from the results of ref.~\cite{gv1}:
\be
 \frac{\partial p^{ }_\rmii{E}}{\partial \nu^2}
  = 
 T^2 \biggl\{ 
  \alpha^{ }_\rmii{E$\nu$}
 + 
 \frac{
   g_1^2 \alpha^{ }_\rmii{EB$\nu$}
 + g_2^2 \alpha^{ }_\rmii{EA$\nu$}
 + \lambda \alpha^{ }_\rmii{E$\lambda\nu$}
 + h_t^2 \alpha^{ }_\rmii{EY$\nu$}
  }{(4\pi)^2}
 \biggr\}
 % \nn 
 + % & & \; + \, 
 \frac{2 \nu^2 }{(4\pi)^2} 
 \biggl(\frac{1}{\epsilon} +  \alpha^{ }_\rmii{E$\nu\nu$} \biggr)
 + \rmO(g^4) 
 \;. \la{dpE}
\ee
Inserting coefficients from ref.~\cite{gv1} and from \eq\nr{aEnn}, 
this contains $1/\epsilon$-divergences in the terms 
proportional to $g_1^2, g_2^2$ and $\nu^2$:
\ba
 \frac{\partial p^{ }_\rmii{E}}{\partial \nu^2}
 & = & 
 \frac{T^2}{6} 
 \biggl\{
  1 - \frac{1}{(4\pi)^2}
 \biggl[
   \fr32 (g_1^2 + 3 g_2^2) 
   \biggl(\frac{1}{\epsilon} + 3 \ln\frac{\bmu}{4\pi T} + \gammaE + \fr53 + 
   \frac{2\zeta'(-1)}{\zeta(-1)} \biggr) 
 \nn & & \; + \, 
  6 h_t^2 \biggl(
   \ln\frac{\bmu}{8\pi T} + \gammaE 
   \biggr) \; + \; 
  12 \lambda \biggl(
   \ln\frac{\bmu}{4\pi T} + \gammaE 
   \biggr)
 \biggr] 
 \biggr\}
 \nn & +  &
 \frac{4 \nu^2}{(4\pi)^2}
 \biggl( \frac{1}{2\epsilon} +
     \ln\frac{\bmu}{4\pi T} + \gammaE   \biggr) 
 + \rmO(g^4)
 \;. \la{dpEnn}
\ea

The second term of \eq\nr{pdp_split}, 
the contribution from the ``soft'' scales $k\sim gT$, 
originates at $\rmO(g^3)$ from the dependence of the effective mass
parameters on $\nu^2$: 
\be
 \frac{\partial p^{ }_\rmii{M}}{\partial \nu^2}
 = 
 \sum_{i=1}^2 
 \frac{\partial m^2_\rmii{E$i$}}{\partial \nu^2} \, 
 \frac{\partial p^{ }_\rmii{M}}{\partial m^2_\rmii{E$i$}}
 + 
 \frac{\partial m^2_3}{\partial \nu^2} \, 
 \frac{\partial p^{ }_\rmii{M}}{\partial m^2_3}
 + \rmO(g^6) \;. \la{dpM_master}
\ee
The first two terms give contributions that 
can be extracted from~\cite{gv2}: the Debye mass parameters
depend on $\nu^2$ as 
\be
 m^2_\rmii{E1} = g_1^2 \, 
 \biggl[ T^2 \beta'_\rmii{E1} - \frac{\nu^2}{(4\pi)^2}
 \beta'_\rmii{E$\nu$} \biggr] + \ldots
 \;, \quad
 m^2_\rmii{E2} = g_2^2 \, 
 \biggl[ T^2 \beta^{ }_\rmii{E1} - \frac{\nu^2}{(4\pi)^2}
 \beta_\rmii{E$\nu$} \biggr] + \ldots
 \;, \la{mEmasses}
\ee
where $\beta'_\rmii{E1}$ and $\beta^{ }_\rmii{E1}$ 
are as in \eq\nr{mEmasses2}. 
Differentiating the term on the second line of \eq\nr{pEpM} 
and inserting the values of the coefficients from ref.~\cite{gv2} yields
\be
 \sum_{i=1}^2 
 \frac{\partial m^2_\rmii{E$i$}}{\partial \nu^2} \, 
 \frac{\partial p^{ }_\rmii{M}}{\partial m^2_\rmii{E$i$}}
 = 
 - \frac{T^2}{(4\pi)^3}
 \Bigl[ 
   3 g_2^3 \Bigl( \fr56 + \frac{\nF}{3} \Bigr)^{\fr12} 
  + g_1^3 \Bigl( \fr16 + \frac{5\nF}{9} \Bigr)^{\fr12}
 \Bigr] 
 + \rmO(g^4) 
 \;. \la{dpM} 
\ee

The last term of \eq\nr{dpM_master} also contributes, but 
this contribution cannot be 
extracted from ref.~\cite{gv2}. The reason is that if we consider 
${p}^{ }_\rmii{M}$ {\em without} a derivative, 
then contributions involving $m_3^2$ 
are of $\rmO(g^6)$ for $|m_3^2| \lsim g^3 T^2/\pi$. 
However, the derivative of $m_3^2$ is larger than
$m_3^2$ itself:  
$\nu^2 \partial m_3^2 / \partial \nu^2 \sim  \nu^2 \sim  g^2 T^2$.
Therefore terms that are of higher order 
in ${p}^{ }_\rmii{M}$ do contribute to the trace anomaly.

%%%%%%%%%%%%%%%%%%%%%%%%% FIGURE %%%%%%%%%%%%%%%%%%%%%%%%%%%%%%%%%%%%%%%%%
%
\begin{figure}[t]

\hspace*{2.5cm}%
\begin{minipage}[c]{10cm}
\begin{eqnarray*}
% \frac{  \partial m^2_{3} }{ \partial \nu^2 }
 \frac{\partial p^{ }_\rmii{M}}{\partial m_3^2} 
 & = & 
 \ScatA \quad + 
 \ScatB \qquad\;. 
\end{eqnarray*}
\end{minipage}

\vspace*{5mm}

\caption[a]{\small 
Processes representing 
$
  \partial p^{ }_\rmii{M} / \partial m^2_{3} 
% ( \partial m^2_{3}/\partial \nu^2 )
$. The filled blob is $m_3^2\,\phi^\dagger\phi$; 
dashed lines are scalar propagators; 
solid lines are adjoint scalar propagators; 
and wiggly lines are gauge fields.  
All 1-loop and 2-loop diagrams and all other 3-loop diagrams
vanish in dimensional regularization. 
}
\la{fig:pMm3}
\end{figure}
%
%%%%%%%%%%%%%%%%%%%%%%%%%%%%%%%%%%%%%%%%%%%%%%%%%%%%%%%%%%%%%%%%%%%%%%%%%%

The computation of $\partial p^{ }_\rmii{M}/\partial m_3^2$ is simplified
by the fact that since the softest physics is captured by a study of 
$p^{ }_\rmii{G}$, the Higgs and gauge fields can be treated as massless
in the computation of the matching coefficient
$p^{ }_\rmii{M}$. Then most diagrams, in particular
all diagrams which do not contain at least one adjoint scalar propagator, 
vanish in dimensional regularization. The leading non-zero contributions, 
which are the only terms needed at $\rmO(g^3)$, are those 
given in \fig\ref{fig:pMm3}. Reducing the diagrams 
to master integrals and evaluating them with standard techniques 
(cf.\ e.g.\ ref.~\cite{AKR}),  we obtain
\ba
 \frac{\partial m^2_3}{\partial \nu^2} \, 
 \frac{\partial p^{ }_\rmii{M}}{\partial m^2_3}
 = 
 -\frac{T^3}{16(4\pi)^3}
 \biggl[
   \frac{g_2^4}{m^{ }_\rmii{E2}} \biggl(-\frac{2}{\epsilon}
   - 12 \ln \frac{\bmu}{2 m^{ }_\rmii{E2} } + \frac{35}{3} \biggr) 
   + \frac{g_1^4}{m^{ }_\rmii{E1}} 
   + \frac{12 g_1^2 g_2^2}{m^{ }_\rmii{E1} + m^{ }_\rmii{E2}}
 \biggr]
 \;. \la{dpM_dm3} 
\ea

%%%%%%%%%%%%%%%%%%%%%%%%%%% SUBSECTION %%%%%%%%%%%%%%%%%%%%%%%%%%%%%%%%%%%
%
\subsection{Non-perturbative contribution}

The third term of \eq\nr{pdp_split}, 
the contribution from the ``ultrasoft'' 
scales $k\sim g^2 T/\pi$, can be expressed as
\be
 \frac{\partial p^{ }_\rmii{G}}{\partial \nu^2} 
 = 
 \frac{\partial \bar{m}_3^2}{\partial \nu^2}
 \frac{\partial p^{ }_\rmii{G}}{\partial \bar{m}_3^2} 
 + \rmO(g^8)
 \;,
\ee
where $\bar{m}_3^2$ is the Higgs mass parameter within MSM
and the error comes from partial derivatives with respect
to the other effective couplings of MSM. 
The dependence of $\bar{m}_3^2$ on $\nu^2$ is known
up to $\rmO(g^2)$~\cite{generic}, 
% \ba
%  - \frac{\partial \bar{m}_3^2}{\partial \nu^2} 
%  & = &
%  1 + \frac{3}{2(4\pi)^2}
%  \biggl[ 
%    \bigl( g_1^2 + 3 g_2^2 - 8 \lambda \bigr)
%    \ln\Bigl( \frac{\bmu e^{\gammaE}}{4\pi T} \Bigr)
%   - 4 h_t^2 
%    \ln\Bigl( \frac{\bmu e^{\gammaE}}{\pi T} \Bigr)
%  \biggr] 
%  + \rmO(g^4)
%  \;, \la{part1}
% \ea
with the leading term reading
$
 {\partial \bar{m}_3^2} / {\partial \nu^2} = -1 
$.
The condensate can be expressed in dimensional regularization as
\be
 - \frac{\partial p^{ }_\rmii{G}}{\partial \bar{m}_3^2} = 
  \frac{(g_\rmii{M1}^2 + 3 g_\rmii{M2}^2)T}{(4\pi)^2}
 \biggl(  \frac{1}{4\epsilon} + \ln\frac{\bmu}{g_\rmii{M2}^2 } \biggr)
 + \langle \phi^\dagger \phi \rangle^{ }_\rmi{3d}(g_\rmii{M2}^2)
 \;, \la{part2}
\ee
where the argument of 
$
 \langle \phi^\dagger \phi \rangle^{ }_\rmi{3d}
$
indicates the renormalization scale used within MSM. 
As discussed below \eq\nr{range} the condensate is 
parametrically $\lsim g T^2$ in our power counting,  
and therefore we keep the correction of $\rmO(g^2)$
in the coefficient of 
$
 \langle \phi^\dagger \phi \rangle^{ }_\rmi{3d}
$, 
even though we do not keep it in the ultraviolet part where it is
unambiguously of $\rmO(g^4)$. 
For future reference we express the ultraviolet part 
in terms of the couplings 
of the full theory, by inserting\footnote{%
 For $g_\rmii{M2}^2$ corrections are known up to 2-loop
 order~\cite{pg}. 
 The other parameters of MSM read: 
 \ba
  \lambda^{ }_\rmii{M} & = & 
  \lambda^{ }_\rmii{E} - 
  \frac{1}{8\pi}
  \biggl[ 
     \frac{3 g_\rmii{E2}^4}{16 m^{ }_\rmii{E2}}
     + \frac{g_\rmii{E1}^2 g_\rmii{E2}^2}
        {4(m^{ }_\rmii{E1}+m^{ }_\rmii{E2})}
     + \frac{g_\rmii{E1}^4}{16 m^{ }_\rmii{E1}}
  \biggr]
  \;, \\
  \bar{m}_\rmii{3B}^2 & = & 
 m_\rmii{3B}^2 - \frac{1}{16\pi} \bigl( g_\rmii{E1}^2 m^{ }_\rmii{E1} + 
 3  g_\rmii{E2}^2 m^{ }_\rmii{E2} \bigr)
   +  \frac{1}{8(4\pi)^2}
  \biggl[ 
    15 g_\rmii{E2}^4 \biggl( 
       \frac{1}{4\epsilon} + \ln\frac{\bmu}{2 m^{ }_\rmii{E2}}  
     + \fr3{10} \biggl)
  \nn 
  & & \; - \, 
  6 g_\rmii{E1}^2 g_\rmii{E2}^2
  \biggl( \frac{1}{4\epsilon} 
  + \ln\frac{\bmu}{m^{ }_\rmii{E1} + m^{ }_\rmii{E2}} + \fr12 \biggr)
  - g_\rmii{E1}^4 
  \biggl( \frac{1}{4\epsilon} 
  + \ln\frac{\bmu}{2 m^{ }_\rmii{E1}} + \fr12 \biggr)
  \biggr]
  \;.
 \ea
 These are needed if one wants to deduce \eq\nr{dpM_dm3} directly
 from \eq(35) of ref.~\cite{gv1} by taking 
 $m^{ }_\rmii{E1}, m^{ }_\rmii{E2} \gg m^{ }_3$.
 }
\ba
 g^2_\rmii{M2} & = & 
% g^2_\rmii{E}\, 
% \biggl( 1 - \frac{g^2_\rmii{E2}}{24\pi m^{ }_\rmii{E2}}
% [ 1 + 2 \epsilon \ln \frac{\bmu}{2 m^{ }_\rmii{E2}} ] + ... \biggr)
% \; = \; 
 g_2^2 T \, 
 \biggl[  1 - \frac{g^2_2 T }{24\pi m^{ }_\rmii{E2} }
 \biggl( 1 + 2 \epsilon \ln \frac{\bmu}{2 m^{ }_\rmii{E2}} \biggr)
  \biggr]
 + \rmO(g^4) \;, \\
 g^2_\rmii{M1} & = & g_1^2 T + \rmO(g^4)
 \;.  
\ea
Thereby the ultrasoft contribution from $k\sim g^2T/\pi$ becomes
\ba
 \frac{\partial \bar{m}_3^2}{\partial \nu^2}
 \frac{\partial p^{ }_\rmii{G}}{\partial \bar{m}_3^2} 
 &  = &  
 \, 
 \biggl\{
   1 + \frac{3}{2(4\pi)^2}
 \biggl[ 
   \bigl( g_1^2 + 3 g_2^2 - 8 \lambda \bigr)
   \ln\Bigl( \frac{\bmu e^{\gammaE}}{4\pi T} \Bigr)
  - 4 h_t^2 
   \ln\Bigl( \frac{\bmu e^{\gammaE}}{\pi T} \Bigr)
 \biggr]  
 \biggr\}
 \, 
 \langle \phi^\dagger \phi \rangle^{ }_\rmi{3d}(g_\rmii{M2}^2)
 \nn
 & + & \frac{(g_1^2 + 3 g_2^2)T^2}{(4\pi)^2}
 \biggl( \frac{1}{4\epsilon} + \ln\frac{\bmu}{g_\rmii{M2}^2} \biggr)
 \nn 
 & - & 
  \frac{g_2^4 T^3}{2 (4\pi)^3 m^{ }_\rmii{E2}}
 \biggl( 
   \frac{1}{4\epsilon} + \ln\frac{\bmu}{g_\rmii{M2}^2} 
   + \fr12 \ln \frac{\bmu}{2 m^{ }_\rmii{E2}} 
 \biggr) 
 + \rmO(g^4)
 \;. \la{part3}
\ea

%%%%%%%%%%%%%%%%%%%%%%%%%%%%%%%%% FIGURE %%%%%%%%%%%%%%%%%%%%%%%%%%%%%%%%%
\begin{figure}[t]

%\vspace*{-3cm}

\centerline{%
 \epsfysize=7.5cm\epsfbox{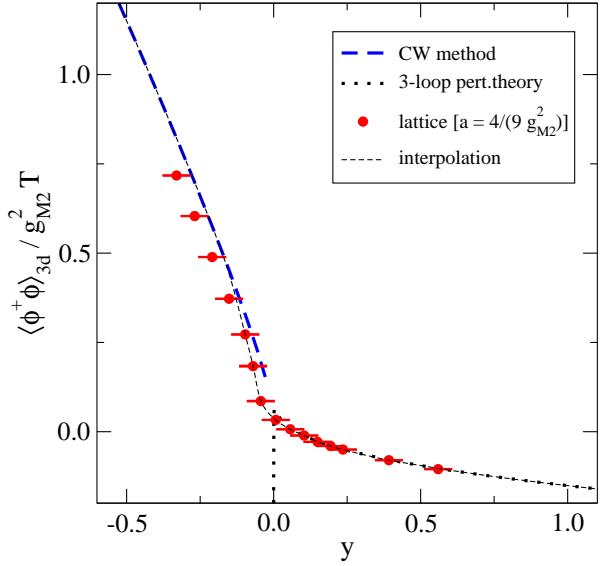}%
% ~~~\epsfysize=7.5cm\epsfbox{imDELTA_M10e7.eps}
}

\caption[a]{\small
 The $\msbar$ renormalized
 3d condensate (cf.\ \eq\nr{part2})
 according to lattice simulations~\cite{anders1} 
 and perturbation theory (cf.\ appendix~A), 
 as a function of $y$ from \eq\nr{3dparams}. 
 The lattice data corresponds to a fixed 
 lattice spacing $a$, which induces errors particularly  
 at low temperatures~\cite{anders0}. 
 For the results of 
 \se\ref{se:all} the phenomenological 
 interpolation indicated with the thin dashed line
 is employed. 
}

\la{fig:condensate}
\end{figure}
%%%%%%%%%%%%%%%%%%%%%%%%%%%%%%%%%%%%%%%%%%%%%%%%%%%%%%%%%%%%%%%%%%%%%%%%%%%

The condensate 
$\langle \phi^\dagger \phi \rangle^{ }_\rmi{3d}(g_\rmii{M2}^2)$
can be measured non-perturbatively by subtracting
proper counterterms~\cite{contlatt} from lattice measurements
extrapolated to the infinite-volume limit, and extrapolating
subsequently to the continuum limit.
A continuum extrapolation has only been carried out at an unphysically
small Higgs mass~\cite{anders0} but cutoff effects have been seen to be
modest, as long as we are not in the broken phase. Therefore 
we make use of lattice results at the physical Higgs mass~\cite{anders1} 
only for $-0.1 \lsim y \lsim 0.2$ 
in terms of the parameters in \eq\nr{3dparams}. 
In order to extrapolate to higher or lower temperatures, 
perturbative expressions are employed; their values are 
discussed in appendix~A. The procedure
is illustrated in \fig\ref{fig:condensate}.

%%%%%%%%%%%%%%%%%%%%%%%%%%% SUBSECTION %%%%%%%%%%%%%%%%%%%%%%%%%%%%%%%%%%%
%
\subsection{Combined expression}

Summing together the contributions from \eqs\nr{dpEnn},
\nr{dpM}, \nr{dpM_dm3} and \nr{part3}, most of the 
$1/\epsilon$-divergences cancel, and we get 
\ba
 - \frac{2 \nu^2 \mathcal{Z}_m \langle \phi^\dagger\phi \rangle }
 {T^4} & = & 
 \Delta^{ }_2(T;\bmu) 
 - \frac{4\nu^4}{(4\pi)^2T^4\epsilon} 
 + \rmO(g^6)
 \;, \la{D2_bare} 
\ea 
where the finite part reads
\ba
 \Delta^{ }_2(T;\bmu) \!\! & \equiv & \!\! 
 -\frac{2 \nu^2 }{T^4}
 \, 
  \biggl\{
    1 + \frac{3}{2(4\pi)^2}
  \biggl[ 
    \bigl( g_1^2 + 3 g_2^2 - 8 \lambda \bigr)
    \ln\Bigl( \frac{\bmu e^{\gammaE}}{4\pi T} \Bigr)
   - 4 h_t^2 
    \ln\Bigl( \frac{\bmu e^{\gammaE}}{\pi T} \Bigr)
  \biggr]  
  \biggr\}
 \, 
 \langle 
   \phi^\dagger\phi 
 \rangle^{ }_\rmi{3d}(g^2_\rmii{M2})
 \nn & - & \!\!
 \frac{\nu^2 }{3 T^2} 
 \biggl\{
  1 - \frac{3}{2(4\pi)^2}
 \biggl[
   (g_1^2 + 3 g_2^2)  
   \biggl(4 \ln\frac{g^2_\rmii{M2}}{\bmu} + 
  3 \ln\frac{\bmu}{4\pi T} + \gammaE + \fr53 + 
   \frac{2\zeta'(-1)}{\zeta(-1)} \biggr) 
 \nn & & \; + \, 
  4 h_t^2 
   \ln \Bigl(\frac{\bmu e^{\gammaE}}{8\pi T} 
   \Bigr) \; + \; 
  8 \lambda 
   \ln \Bigl(\frac{\bmu e^{\gammaE}}{4\pi T} 
   \Bigr)
 \biggr] 
 \biggr\}
 \nn & + & \!\! 
   \frac{2 \nu^2 }{(4\pi)^3 T^2}
 \biggl[
   \frac{g_1^2 m^{ }_\rmii{E1} + 3 g_2^2 m^{ }_\rmii{E2} }{T}
  % \Bigl( \fr56 + \frac{\nF}{3} \Bigr)^{\fr12} 
  % \Bigl( \fr16 + \frac{5\nF}{9} \Bigr)^{\fr12}
  + \frac{g_1^4 T}{16 m^{ }_\rmii{E1}}
  + \frac{3 g_1^2 g_2^2 T}{4(m^{ }_\rmii{E1}+m^{ }_\rmii{E2})}
  + %% \nn & & \; + \, 
  \frac{g_2^4 T}{2 m^{ }_\rmii{E2}}
  \biggl( \frac{35}{24}  + \ln \frac{2 m^{ }_\rmii{E2}}{g_\rmii{M2}^2 }
  \biggr)
 \biggr] 
 \nn & - & \!\!
 \frac{8 \nu^4}{(4\pi)^2 T^4}
    %% \frac{1}{2\epsilon} +
     \ln \Bigl( \frac{\bmu  e^{\gammaE}}{4\pi T}   \Bigr) 
 \;. \la{t3}
\ea
Apart from the last term the result is 
$\bmu$-independent up to $\rmO(g^6)$. 
This $\bmu$-dependence as well as the divergence in 
\eq\nr{D2_bare} cancel against terms from the 
vacuum subtraction, cf.\ \se\ref{se:vac}.

%%%%%%%%%%%%%%%%%%%%%%%%%%% SECTION %%%%%%%%%%%%%%%%%%%%%%%%%%%%%%%%%%%%%%
%
\section{Vacuum subtraction and renormalization}
\la{se:vac}

Suppose that we compute the bare vacuum pressure, $p^{ }_\rmii{0B}$, 
in a perturbative loop expansion: 
$p^{ }_\rmii{0B} = \sum_{\ell} p_\rmii{0B}^{(\ell)}$. 
In a gauge-fixed computation, the result is 
a function of the Higgs expectation value, which can likewise
be determined order by order: $v = \sum_{\ell} v^{(\ell)}$. 
Even though $v$ is gauge-dependent, $p^{ }_\rmii{0B}$ is gauge-independent
order by order in perturbation theory. Inserting $v$ into $p^{ }_\rmii{0B}$, 
we can re-expand the result as 
\be
 p^{ }_\rmii{0B} = p_\rmii{0B}^{(0)}\Bigl(v^{(0)} + v^{(1)} + ...\Bigr) + 
 p_\rmii{0B}^{(1)}\Bigl(v^{(0)} + ...\Bigr) + ... 
 = 
 p_\rmii{0B}^{(0)}\Bigl(v^{(0)}\Bigr) 
 + p_\rmii{0B}^{(1)}\Bigl(v^{(0)}\Bigr) + ... 
 \;, 
\ee
where we made use of $p_\rmii{0B}^{(0)} = p_\rmii{0R}^{(0)}$ and
$[p_\rmii{0R}^{(0)}]'(v^{(0)}) = 0$. 
The terms not shown are of $\rmO(g^6)$
in our power counting. 
Therefore, it is sufficient for our purposes to compute $p^{ }_\rmii{0B}$
up to 1-loop level and insert the tree-level Higgs vacuum 
expectation value  $v^2 = \nu^2/\lambda$
into the expression. 

Dropping the superscripts, we get 
\ba
 p^{ }_\rmii{0B} & = & \biggl\{  \fr12 ( \nu^2 + \delta \nu^2 ) v^2 
 - \fr14 (\lambda + \delta \lambda) v^4 
 \nn & & \; + \,
 \frac{3 \mW^4}{32\pi^2} 
 \biggl( \frac{1}{\epsilon} + \ln \frac{\bmu^2}{\mW^2} + \fr56 \biggr)
 + 
 \frac{3 \mZ^4}{64\pi^2} 
 \biggl( \frac{1}{\epsilon} + \ln \frac{\bmu^2}{\mZ^2} + \fr56 \biggr)
 \nn & & \; + \, 
 \frac{\mH^4}{64\pi^2}
 \biggl( \frac{1}{\epsilon} + \ln \frac{\bmu^2}{\mH^2} + \fr32 \biggr)
 -  
 \frac{3 m_t^4}{16\pi^2}
 \biggl( \frac{1}{\epsilon} + \ln \frac{\bmu^2}{m_t^2} + \fr32 \biggr)
 \biggr\}_{v^2 = \nu^2/\lambda}
 + \rmO(g^6)
 \;.
\ea
The counterterms read
\ba
 \delta \nu^2 & = & \frac{3\nu^2}{(4\pi)^2\epsilon}
 \biggl[
   -\frac{g_1^2 + 3 g_2^2}{4} + h_t^2 + 2 \lambda 
 \biggr]
 \;, \\ 
 \delta \lambda & = & \frac{3}{(4\pi)^2\epsilon}
 \biggl[
   \frac{g_1^4 + 2 g_1^2 g_2^2 + 3 g_2^4}{16}
    - h_t^2 \bigl( h_t^2 - 2 \lambda \bigr) +
    \lambda \Bigl( 4 \lambda - \frac{g_1^2 + 3 g_2^2}{2} \Bigr) 
 \biggr]
 \;,
\ea
whereas at the minimum the masses take the values 
\be
 \mW^2 = \frac{g_2^2\nu^2}{4\lambda} \;, \quad
 \mZ^2 = \frac{(g_1^2 + g_2^2)\nu^2}{4\lambda} \;, \quad
 \mH^2 = 2 \nu^2 \;, \quad
 m_t^2 = \frac{h_t^2\nu^2}{2\lambda} \;.
\ee
Most divergences cancel, 
and the vacuum result becomes
\be
 \frac{4 p^{ }_\rmii{0B}}{T^4} = 
 \Delta^{ }_3(T;\bmu) + \frac{4\nu^4}{(4\pi)^2 T^4 \epsilon}
 \;, \la{Del3}
\ee
where
\ba
 \Delta^{ }_3(T;\bmu) & \equiv &
 \frac{\nu^4}{\lambda T^4} + \frac{4 \nu^4}{(4\pi)^2 T^4}
 \biggl[  \ln\frac{\bmu^2}{\nu^2} + \fr32 \biggr]
 \nn & & \; + \, 
 \frac{3 \nu^4}{\lambda^2 (16\pi)^2 T^4}
 \biggl\{
   2 g_2^4 \biggl[ \ln\frac{4\lambda\bmu^2}{g_2^2 \nu^2} + \fr56 \biggr]
  + 
  (g_1^2 + g_2^2)^2
  \biggl[ \ln\frac{4\lambda\bmu^2}{(g_1^2 + g_2^2) \nu^2} + \fr56 \biggr] 
 \biggr\}
 \nn & & \; - \, 
 \frac{3 \nu^4 h_t^4}{\lambda^2 (4\pi)^2 T^4}
 \biggl[ 
   \ln\frac{2\lambda\bmu^2}{h_t^2 \nu^2} + \fr32
 \biggr]
 + \rmO(g^6)
 \;. \la{t4}
\ea

In practice, it is not useful to evaluate \eq\nr{t4} directly, 
because our renormalization scale will be $\bmu\sim \pi T$, which 
would introduce large logarithms if inserted into \eq\nr{t4}. Rather, 
all vacuum parameters are first evaluated at a scale $\bmu^{ }_0 \equiv \mZ$, 
and then evolved from $\bmu^{ }_0$ to the thermal scale through RG
equations. As an illustration, the 1-loop RG equations for the two 
dimensionful parameters appearing in our analysis read
\ba
 \bmu \frac{{\rm d}\nu^2}{{\rm d}\bmu} & = &  
 \frac{3 \nu^2}{8\pi^2}
 \biggl( 
   - \frac{g_1^2 + 3 g_2^2}{4} + h_t^2 + 2\lambda
 \biggr)
 \;, \la{rg1} \\
 \bmu \frac{{\rm d} p^{ }_\rmii{0R}}{{\rm d}\bmu} & = & 
 \frac{\nu^4}{8 \pi^2}
 \;, \la{rg2}
\ea
from where we get 
$
 \Delta^{ }_3(T;\bmu) = 4 p^{ }_\rmii{0R} / T^4
$.
At the scale $\bmu = \bmu^{ }_0$, the running
couplings are expressed in terms of physical parameters through
1-loop relations specified e.g.\ in ref.~\cite{generic}. 
(For many parameters 
a higher loop order could be extracted from more recent
literature, but such corrections are smaller than thermal uncertainties, 
so for the sake of simplicity
and reproducibility we make use of explicit 1-loop expressions.)

%%%%%%%%%%%%%%%%%%%%%%%%%%% SECTION %%%%%%%%%%%%%%%%%%%%%%%%%%%%%%%%%%%%%%
%
\section{Phenomenological results}
\la{se:all}

Summing together the terms from \eqs\nr{t12}, \nr{D2_bare}, \nr{Del3} as
dictated by \eq\nr{master},  all $1/\epsilon$-divergences cancel,  
and we obtain 
\ba
%  & & \hspace*{-1cm}
  T \frac{{\rm d}}{{\rm d}T}
 \biggl\{ \frac{p^{ }_{ }(T)}{T^4} \biggr\}
%%%
 & = &  
 \Delta^{ }_1(T) \; + \; 
 \Delta^{ }_2(T;\bmu) \; + \; 
 \Delta^{ }_3(T;\bmu) \; + \; 
 \rmO(g^6)
 \;. \la{trace}
\ea
Because of a cancellation between $\Delta^{ }_2$ and $\Delta^{ }_3$
the result is formally
independent of the renormalization scale $\bmu$, even though in practice a 
residual $\bmu$-dependence is left over, as a reflection of unknown
higher order corrections. We write  
$\bmu = \alpha \pi T$, and vary
$\alpha$ in the range $\alpha\in (0.5 ... 2.0)$ in order to get
one impression on the corresponding uncertainty. 

%%%%%%%%%%%%%%%%%%%%%%%%%%%%%%%%% FIGURE %%%%%%%%%%%%%%%%%%%%%%%%%%%%%%%%%
\begin{figure}[t]

%\vspace*{-3cm}

\centerline{%
 \epsfysize=7.5cm\epsfbox{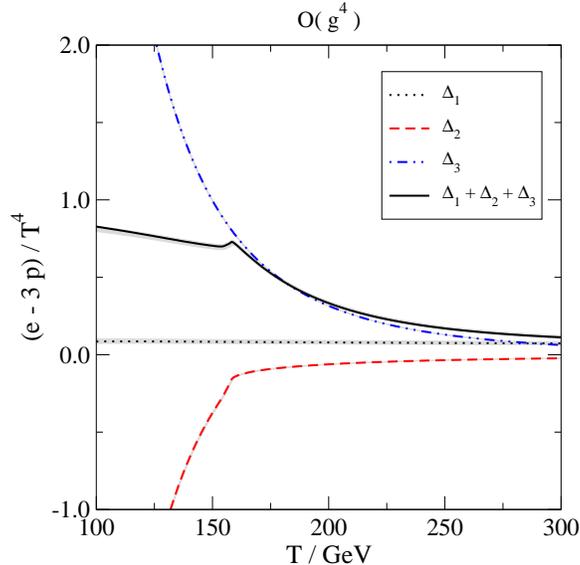}%
% ~~~\epsfysize=7.5cm\epsfbox{imDELTA_M10e7.eps}
}

\caption[a]{\small
 The trace ``anomaly'' from \eq\nr{trace}. The grey band reflects variations
 of the renormalization scale in the range
 $
  \bmu = (0.5 ... 2.0) \pi T
 $. 
 Contributions of $\rmO(\nu^4/T^4)$ are seen to cancel 
 at low temperatures.  
}

\la{fig:trace}
\end{figure}
%%%%%%%%%%%%%%%%%%%%%%%%%%%%%%%%%%%%%%%%%%%%%%%%%%%%%%%%%%%%%%%%%%%%%%%%%%%

The result of \eq\nr{trace} is accurate up to and including
the order $\rmO(g^5)$. It is well known from studies of the pressure of 
QCD, however, that certain odd orders show anomalously poor convergence. 
In particular, whereas the $\rmO(g^2)$ correction to the pressure provides
for a reasonable approximation, the $\rmO(g^3)$ correction is far off. 
For $\Delta$, the order $\rmO(g^4)$ is related to the 
$\rmO(g^2)$ correction to the pressure, 
and the order $\rmO(g^5)$ to the $\rmO(g^3)$ correction 
(cf.\ \se\ref{se:quantum}).
For this reason, the numerically most accurate estimate for $\Delta$
can probably be obtained by restricting to $\rmO(g^4)$.
A result corresponding to this accuracy is 
shown for the observable 
of \eq\nr{trace} in \fig\ref{fig:trace}. 
(For the pressure we display also the $\rmO(g^5)$ result
in \fig\ref{fig:phen}(left).) 

In order to obtain other thermodynamic functions, the boundary value 
needed for \eq\nr{integral_method} should be fixed. We do this 
on the low-temperature side, making use of the 
results of ref.~\cite{pheneos}\footnote{%
 Tabulated results can be downloaded from \la{fn:highT}
 {\tt www.laine.itp.unibe.ch/eos06/}.
 }
and thereby setting 
$p(T^{ }_0) / T_0^4 \simeq 10.91$ at $T^{ }_0 = 100$~GeV. 
On the high-temperature side, we expect to match to the results of 
ref.~\cite{gv1}, after changing the overall renormalization condition
to our \eq\nr{renorm}
and by making the changes listed in appendix~B. Like for $\Delta$, 
we make use of the result of $\rmO(g^4)$, 
with the negative $\rmO(g^5)$ QCD contribution
expected to lead to an underestimate~\cite{gsixg,buwu}.
(We have also experimented with an 
approximate $\rmO(g^6)$ QCD contribution as
estimated in ref.~\cite{pheneos}, finding a result which is in between
the $\rmO(g^4)$ and $\rmO(g^5)$ ones, confirming that the
$\rmO(g^5)$ result is most likely on the low side.)

After choosing an initial condition, 
other thermodynamic functions are obtained as follows: 
the pressure $p^{ }_{ }/T^4$ from \eq\nr{integral_method}; 
the energy density from 
$
 e^{ }_{ }/T^4 = \Delta + 3 p^{ }_{ }/T^4
$; 
the entropy density $s^{ }_{ } = p'_{ }$ from 
$
 s^{ }_{ }/T^3 = \Delta + 4 p^{ }_{ }/T^4
$;  
the heat capacity $c^{ }_{ } = e'_{ }$ from
$
 c^{ }_{ }/T^3 = T \Delta'
 + 7 \Delta + 12 p^{ }_{ }/T^4
$; 
the equation-of-state parameter from 
$
 w^{ }_{ } = p^{ }_{ } / e^{ }_{ } = 
 1 / (3 + \Delta T^4 / p^{ }_{ })
$; 
and the speed of sound squared from 
$
 c_s^2 = p'_{ } / e'_{ } = s^{ }_{ }/ c^{ }_{ }  
$.
Some of these functions are conveniently parametrized through
\ba
 g_\rmi{eff}(T)  \equiv
   \frac{e^{ }_{ }(T)}{\Bigl[\frac{\pi^2 T^4}{30}\Bigr]}
 \;, \quad
 h_\rmi{eff}(T)  \equiv 
  \frac{s^{ }_{ }(T)}{\Bigl[\frac{2\pi^2 T^3}{45}\Bigr]}
 \;, \quad
 i_\rmi{eff}(T)  \equiv
   \frac{c^{ }_{ }(T)}{\Bigl[\frac{2\pi^2 T^3}{15}\Bigr]}
 \;. \la{ieff} 
\ea
Results for all of these functions are shown in \fig\ref{fig:phen}. 
(The heat capacity and the speed of sound squared are the most
``difficult'' quantities, because they require taking a numerical
temperature derivative from $\Delta$.)

%%%%%%%%%%%%%%%%%%%%%%%%%%%%%%%%% FIGURE %%%%%%%%%%%%%%%%%%%%%%%%%%%%%%%%%
\begin{figure}[t]

\centerline{%
\epsfysize=4.8cm\epsfbox{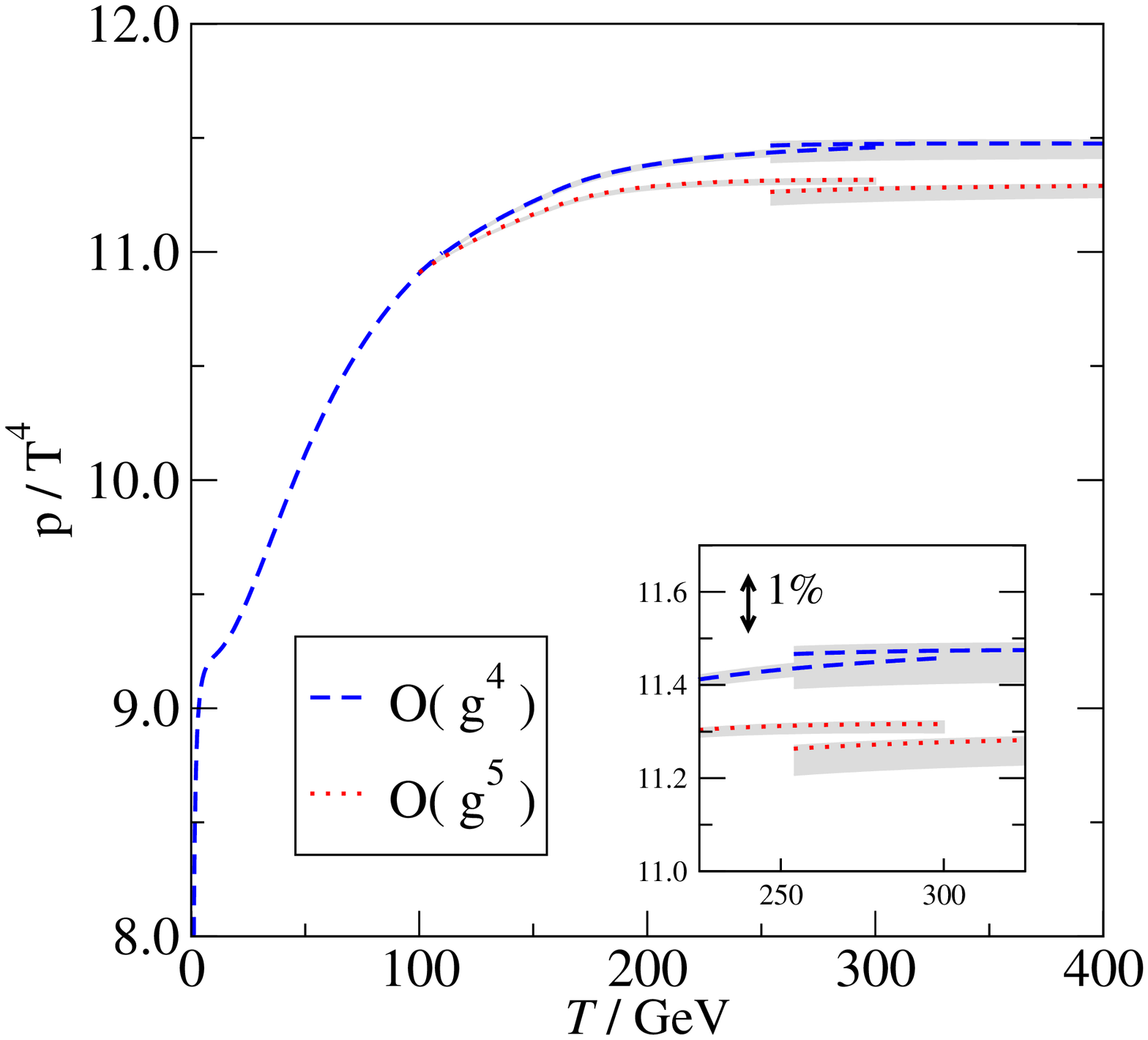}%
~~\epsfysize=5.1cm\epsfbox{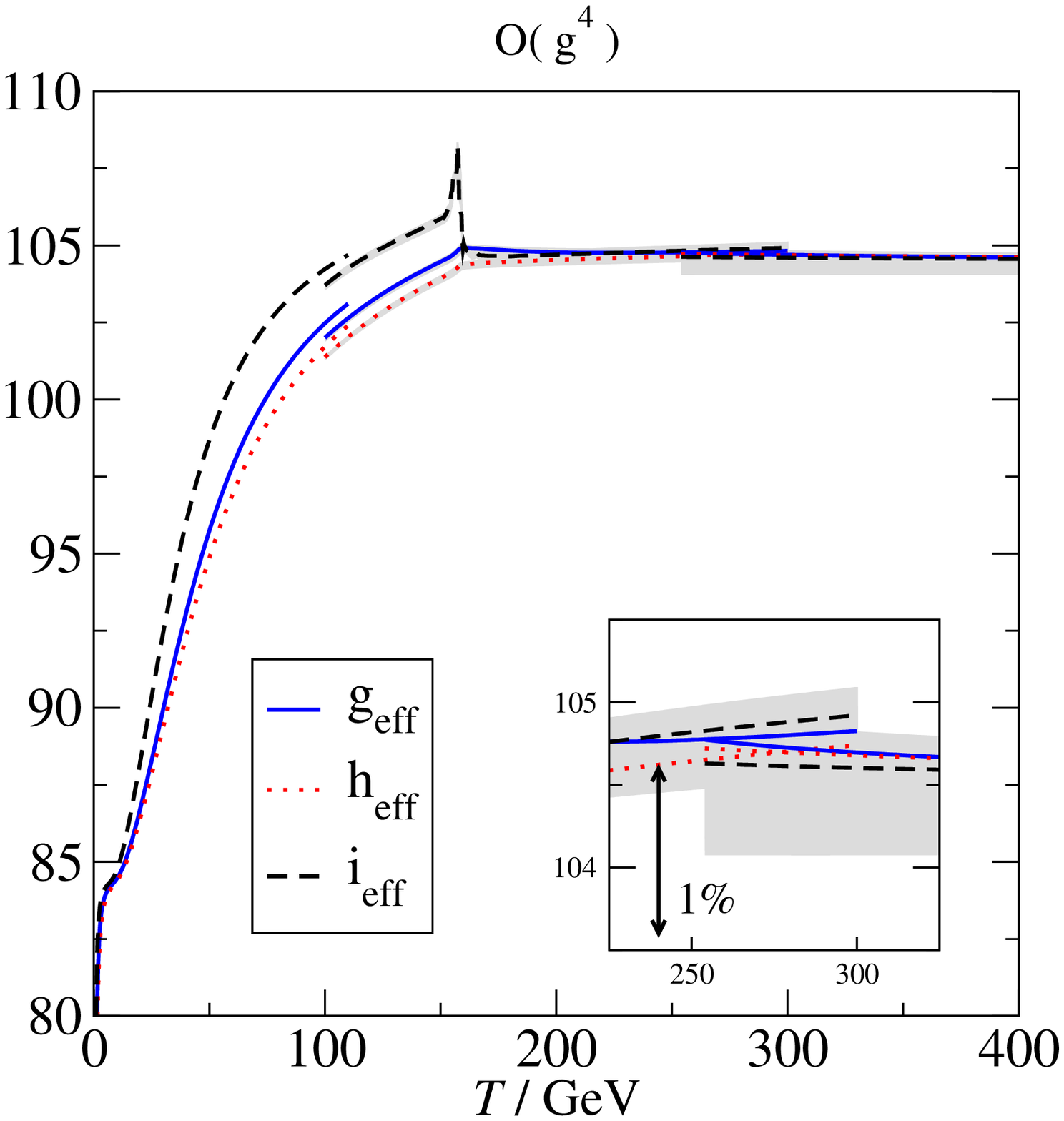}%
~~\epsfysize=5.1cm\epsfbox{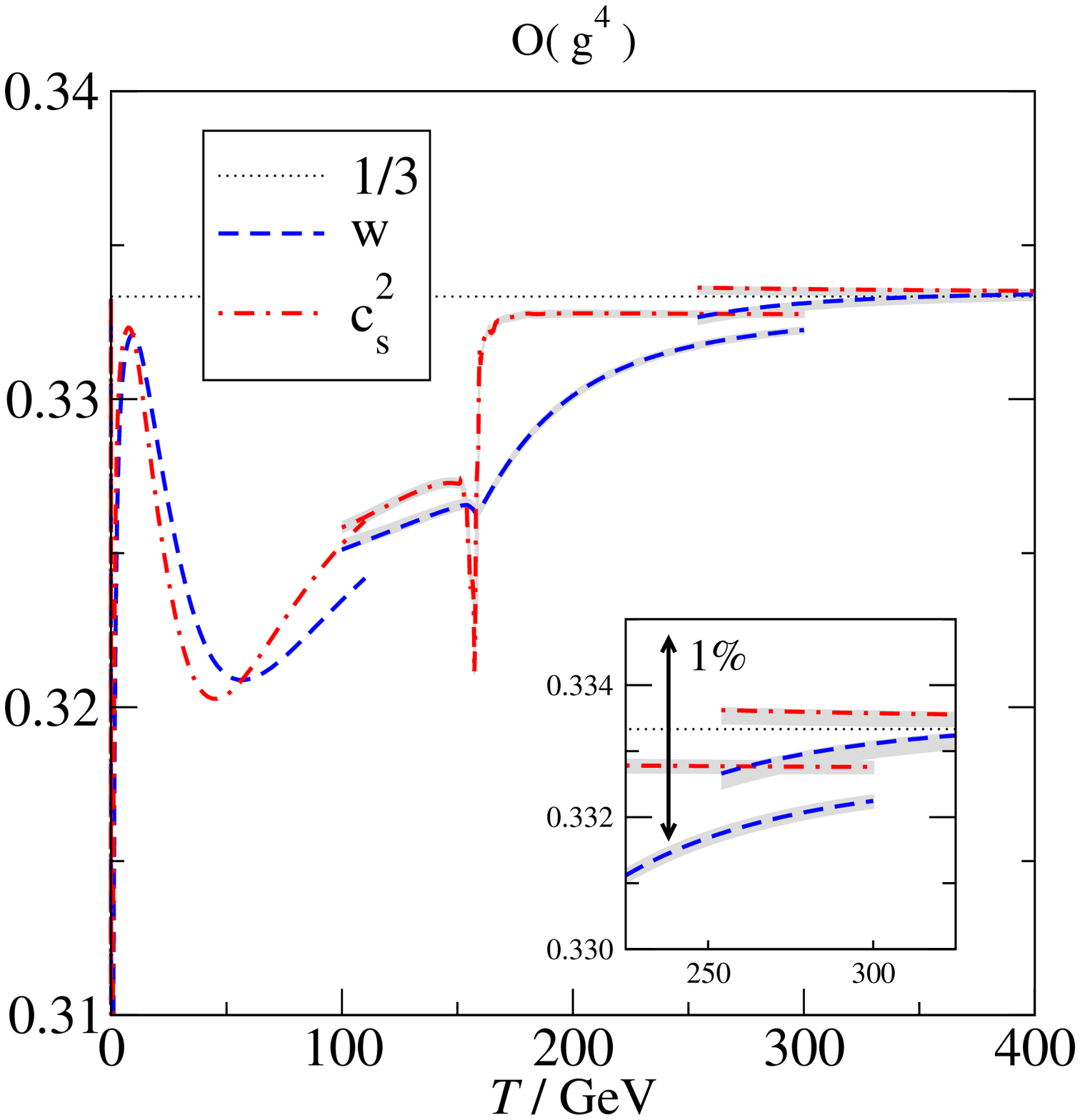}%
}

\caption[a]{%% \small
 Left: The Standard Model pressure. 
 Middle: $g_\rmi{eff}, h_\rmi{eff}, i_\rmi{eff}$ as defined 
 in~\eq\nr{ieff}. The heat capacity (parametrized by $i_\rmi{eff}$)
 shows a narrow peak as is characteristic of a rapid crossover.  
 Right: the equation-of-state parameter $w$ 
 and the speed of sound squared $c_s^2$.
 The grey bands reflect variations
 of the renormalization scale in the range
 $
  \bmu = (0.5 ... 2.0) \pi T
 $. 
 The low and high-$T$ results correspond to ref.~\cite{pheneos} 
($T \lsim 110$~GeV)
 and ref.~\cite{gv1} ($T \gsim 250$~GeV).
 The close-ups illustrate our estimates of theoretical 
 uncertainties on the high-$T$ side. 
} 

\la{fig:phen}
\end{figure}
%%%%%%%%%%%%%%%%%%%%%%%%%%%%%%%%%%%%%%%%%%%%%%%%%%%%%%%%%%%%%%%%%%%%%%%%%%%

It may be wondered why our new results 
do not match exactly the previous ones for high temperatures~\cite{gv1}, 
even though both have been computed up to the same order in $g$. 
The main reason
is that they involve different sets of higher-order corrections. 
In particular, \eq\nr{trace} has been evaluated 
``strictly'' to $\rmO(g^4)$ or $\rmO(g^5)$, because this 
has led to fairly simple expressions. In contrast, 
the high-temperature result is more cumbersome, 
including more terms (with non-zero
corrections of $\rmO(1)$, $\rmO(g^2)$, $\rmO(g^3)$) and 
complicated ``soft'' contributions, because the thermal mass
parameters associated with the Higgs and gauge fields
are of the same order. 
For evaluating such contributions the experience from
QCD suggests that it is not worth re-expanding them 
in terms of the original couplings  
but rather to evaluate $p^{ }_\rmii{M}$ ``as is''. 
We have followed separate procedures for the two regimes, 
because their difference permits for us to get another 
impression on the magnitude of unknown higher-order corrections. 

Among the various types of error estimates that we have made, that 
based on scale variations is clearly a lower bound, because it only 
probes the magnitude of special types of corrections. The error 
estimate originating from the mismatches of various computations 
should in principle be a more reliable one, because all types of 
corrections are
included; nevertheless it should still be treated as a lower bound. 
In practice, depending on the observable, one of the two gives 
a more conservative error estimate.  In \fig\ref{fig:phen} both
estimates are shown; consequently  
we expect the theoretical uncertainty of our analysis to be 
on the percent level. Ultimately, the true accuracy can only 
be judged by carrying out a non-perturbative analysis 
of the observables that we have considered.  

With the help of the functions in \eq\nr{ieff}, 
the relationship of time and temperature in the Early
Universe (assuming a flat geometry)
can be expressed as 
\be
 \fr32 \sqrt{\frac{5}{\pi^3}}
 \frac{m_\rmi{Pl}}{T^3}
 \frac{{\rm d}T}{{\rm d}t} = - 
 \frac{\sqrt{g_\rmi{eff}(T)} h_\rmi{eff}(T)}{i_\rmi{eff}(T)}
 \;.
\ee 
It is seen that a peak in heat capacity
(i.e.\ $i_\rmi{eff}$), visible in \fig\ref{fig:phen}(middle),  
leads to a short period of slower temperature change~\cite{ikkl}, 
and correspondingly
a mildly increased abundance of produced particles if a particle
production process is under way, or a reduced density
of weakly interacting relics  
disappearing through a co-annihilation process.

%%%%%%%%%%%%%%%%%%%%%%%%%%% SECTION %%%%%%%%%%%%%%%%%%%%%%%%%%%%%%%%%%%%%%
%
\section{Conclusions}
\la{se:concl}

Drawing on existing multiloop computations~\cite{gv1} 
and lattice simulations within a dimensionally reduced effective 
theory~\cite{anders1}, and complementing these through new 
1-loop, 2-loop and 
3-loop computations needed for determining the ``trace anomaly'' of the 
electroweak theory, we have estimated the basic thermodynamic
functions that play a role for Standard Model thermodynamics
at temperatures between 100~GeV and
300~GeV (cf.\ \fig\ref{fig:phen}).
These results can be matched, for most observables 
with modest ($\sim 1\%$) discrepancies, to 
perturbative computations at low~\cite{pheneos} or high~\cite{gv1}
temperatures.

One finding of our study is that despite the high temperatures 
considered, radiative corrections are larger than might 
naively be anticipated. Larger corrections also imply that
uncertainties are less well under control. 
On the low-temperature side, we believe that the results of 
ref.~\cite{pheneos} do contain uncertainties because
the analysis of the 
electroweak sector was based on a low loop order and because 
QCD corrections, whose convergence is slow at finite
temperatures, are substantial. On the high-temperature side, 
the QCD corrections continue to be an issue (cf.\ \fig\ref{fig:phen}(left)).
As an example of the physical significance of these uncertainties, it may
be noted from \fig\ref{fig:phen}  that 
the effective numbers of degrees of freedom 
remain below the canonical value 106.75 in 
the whole temperature range considered, and even 
decrease modestly as the temperature 
increases above 300~GeV.   
Even though there could be valid physics reasons for the decrease
(such as that the effective Higgs mass parameter is very small across
the crossover but then increases again), a similar tendency also originates
from the $\rmO(g^4)$ QCD contribution and is then an artifact of the
truncation. (The $\rmO(g^5)$ correction decreases the effective numbers 
of degrees of freedom further but turns the results into slowly
increasing functions).
Another peculiar feature of the $\rmO(g^4)$ result is that 
the speed of sound squared
$c_s^2$ is $\sim 0.1$\% {\it above} $1/3$ at $T > 300$~GeV, 
however this is again reversed by the $\rmO(g^5)$ correction.  
In general, the estimated theoretical uncertainty of our
results is on the percent level, and no conclusions should be drawn
from features finer than this. 

In comparison with ref.~\cite{pheneos}, 
we find results for the effective numbers 
of degrees of freedom that are about 1\% lower at $T > 300$~GeV. 
The reason is mostly due to negative electroweak radiative corrections
proportional to $g_2^2$ and $h_t^2$, which were omitted
in ref.~\cite{pheneos}.  

Apart from radiative corrections, 
another important ingredient in our results is the 3d condensate 
$
  \langle \phi^\dagger \phi \rangle^{ }_\rmi{3d}
$ (cf.\ \eq\nr{part3}), 
measured non-perturbatively on the lattice and subsequently 
converted to the $\msbar$ scheme. It is interesting to note 
that apart from the known strong correlation of 
$
  \langle \phi^\dagger \phi \rangle^{ }_\rmi{3d}
$
with the anomalous rate of baryon plus lepton 
number violation (cf.\ refs.~\cite{krs,anders1} 
and references therein), 
the same quantity also plays a role for other cosmologically
relevant observables, such as the production rate of non-relativistic
right-handed neutrinos~\cite{nonrel} or the relationship between lepton
and baryon number densities~\cite{sangel}. Therefore, it seems well 
motivated to improve on the existing measurements~\cite{anders1}
by including the U(1) subgroup and by taking the continuum limit
for a physical Higgs mass. In addition, it would be helpful to 
measure the susceptibility related to this condensate, 
so that taking a numerical temperature derivative, needed for estimating
the heat capacity or the speed of sound squared, could be avoided. 

Our interpolations of all thermodynamic functions
shown in \fig\ref{fig:phen} 
can be downloaded, for a wide temperature interval, from 
{\tt www.laine.itp.unibe.ch/eos15/}.
(In these results we switch from one regime 
to another with a temperature gap of 15--20~GeV in between.)
Despite the remaining uncertainties, 
we hope that these results turn out to be helpful as a background equation of
state in Dark Matter or Leptogenesis computations operating in 
this temperature range. 

%%%%%%%%%%%%%%%%%%%%%%%%% SECTION %%%%%%%%%%%%%%%%%%%%%%%%%%%%%%%%%%%%%
%
\section*{Acknowledgements}

We thank M.~D'Onofrio, K.~Rummukainen and A.~Tranberg for providing
us with numerical data from ref.~\cite{anders1}.
This work was partly supported by the Swiss National Science Foundation
(SNF) under grant 200021-146737.

%%%%%%%%%%%%%%%%%%%%%%% APPENDIX %%%%%%%%%%%%%%%%%%%%%%%%%%%%%%%%%%%
%
\appendix
\renewcommand{\thesection}{Appendix~\Alph{section}}
\renewcommand{\thesubsection}{\Alph{section}.\arabic{subsection}}
\renewcommand{\theequation}{\Alph{section}.\arabic{equation}}

%%%%%%%%%%%%%%%%%%%%%%%%%%%%%% SECTION %%%%%%%%%%%%%%%%%%%%%%%%%%%%%%%%%
%
\section{3d condensate in perturbation theory}
\la{app:A}

We collect here results for the $\msbar$ renormalized condensate
$
 \langle \phi^\dagger\phi \rangle^{ }_\rmi{3d}
$
defined within the MSM effective theory 
and evaluated at the scale $\bmu = g^2_\rmii{M2}$
(cf.\ \eq\nr{part2}).

For $\bar{m}_3^2 \gg 0$, a 3-loop perturbative expression 
for 
$
  \langle \phi^\dagger \phi \rangle^{ }_\rmi{3d}
$
can be extracted from  \eq(35) of ref.~\cite{gv1}, after subtracting 
the contributions containing adjoint scalar fields
(the latter need to be corrected as explained in appendix~B). 
The expression reads
\ba
 && \hspace*{-1.5cm}
  \frac{
  \langle \phi^\dagger\phi \rangle^{ }_\rmi{3d}(g_\rmii{M2}^2)
  }{
   g_\rmii{M2}^2 T
  }
   \; = \;  
  -\frac{\sqrt{y} }{2\pi} +
  \frac{1}{(4\pi)^2}
  \biggl[ 6 x 
  + \, (3 + z)
  \biggl(
  - \frac{ \ln y}{2} - \ln 2  + \fr14 \biggr)
 \biggr]
 \\  \la{pdp_np_sym} 
%%%%%%%%
 & + &
 \frac{1}{(4\pi)^3\sqrt{y}}\biggl[ 
   \frac{51 \ln y}{32} +\frac{61\ln 2}{16} +\frac{3\pi^2}{16} +\frac{485}{64}
 \, + \, 
 x\, \biggl( 
   \frac{9 \ln y}{2} + 3 \ln 2 +\frac{39}{4}
 \biggr)
%%%%%%%%
 \nn  & + & 
 x^2\, \biggl( 
   -6 \ln y -24 \ln 2 +\frac{3}{2}
 \biggr)
 \, + \, 
 z\, \biggl( 
   -\frac{9 \ln y}{16} - \frac{27\ln 2}{8} +\frac{\pi^2}{8} +\frac{51}{32}
 \biggr)
%%%%%%%%
 \nn  & + & 
 z^2\, \biggl( 
   -\frac{5 \ln y}{32} - \frac{41\ln 2}{48} +\frac{\pi^2}{48} +\frac{47}{192}
 \biggr)
 \, + \, 
 x\, z\, \biggl( 
   \frac{3 \ln y}{2} - 3 \ln 2 +\frac{21}{4}
 \biggr)
 \biggr]
 + \rmO\Bigl( \frac{1}{ y }\Bigr)
  \;, \nonumber
\ea
where
\ba
 x & \equiv & \frac{\lambda^{ }_\rmii{M}}{g_\rmii{M2}^2}
 \;, \quad
 y \; \equiv \; \frac{\bar{m}_3^2(g_\rmii{M2}^2)}{g_\rmii{M2}^4}
 \;, \quad
 z \; \equiv \; \frac{g_\rmii{M1}^2}{g_\rmii{M2}^2}
 \;. \la{3dparams}
\ea
%
% where the mass parameters read
% \ba
%   m_3^2 & = &   -\nu^2  + 
%  \biggl(\frac{\lambda}{2} + \frac{|h_t|^2}{4} + 
%  \frac{g_1^2 + 3 g_2^2}{16} \biggr) T^2
%  -  \frac{( g_1^2 m_\rmii{D1}^{} + 3 g_2^2 m_\rmii{D2}^{ })T}
%   {16\pi } 
%  \;, \\ 
%  m_\rmii{D1}^2 & = &  \frac{11}{6} g_1^2 T^2
%  \;, \quad
%  m_\rmii{D2}^2 \; =  \; \frac{11}{6} g_2^2 T^2
%  \;. \la{mH}
% \ea
As illustrated in \fig\ref{fig:condensate} (dotted lines), the result agrees 
surprisingly well with lattice data as soon as $y \gsim 0.2$. 

In the broken symmetry phase, i.e.\ for $\bar{m}_3^2 \ll 0$, an explicit
loopwise result is available up to 2-loop level, however
it shows poor convergence~\cite{nonpert}. Therefore we make use of 
a numerically determined value 
(referred to as the ``Coleman-Weinberg (CW) method'')
which includes a subset of higher-order corrections and 
has been tested against lattice
simulations in refs.~\cite{nonpert,anders0}. The result
is only available in the approximation $g_\rmii{M1}^{ } = 0$ but
the corrections originating from $g_\rmii{M1}^{ }$ are expected 
to be small. A comparison with lattice data is shown 
in \fig\ref{fig:condensate} (long-dashed line), with the low-temperature
deviation expected to be reduced by a continuum extrapolation~\cite{anders0}. 

% \ba 
%   \frac{
%   \langle \phi^\dagger\phi \rangle^{ }_\rmi{3d}(g_\rmii{M2}^2)
%   }{
%    g_\rmii{M2}^2 T
%   }
% & = & 
% -\frac{y}{2 x}
% + \frac{1}{4\pi}
% \Bigl( 2\sqrt{2 x} + \frac{3}{4 x} \Bigr) \sqrt{-\frac{y}{4x}}
% \nn 
% &  & \; + \,  
% \frac{3}{(16\pi)^2 x}
% \biggl[ 
%    \Bigl(
%     \fr{17}2 + 8 x - 32 x^2 
%     \Bigr) \ln \sqrt{-\frac{y}{4x}}
% \nn 
% & & \; 
% - \, \Bigl(2 - 8 x + 16 x^2 \Bigr) \ln(2+2\sqrt{2x}) + 4\sqrt{2x} - 
% 2x + 8 \sqrt{2} x^{3/2} 
% \nn & & \; - \, 
% 16 x^2 \ln(6 \sqrt{2x}) + \frac{21}{2} \ln 3 + \fr32 + 
% \frac{3}{4x} + 16 x^2 
% \biggr] 
% \;.
%\ea 

%%%%%%%%%%%%%%%%%%%%%%%%%%%%%% SECTION %%%%%%%%%%%%%%%%%%%%%%%%%%%%%%%%%
%
\section{Differences with respect to ref.~\cite{gv1}}
\la{app:B}

Our results make use of expressions first 
derived in ref.~\cite{gv1}, however there are
a few technical points on which we disagree with this reference.
As already discussed in the text, 
an important issue is that of the overall renormalization
condition, \eq\nr{renorm}, which was imposed in a different form 
in ref.~\cite{gv1} (what was subtracted there was the pressure
of a hypothetical zero-temperature ``symmetric phase'').

In addition, 
 the following typographic errors have been detected in ref.~\cite{gv1}: 
 in \eq(8), $ -3 g_Y^4 \to - \Nc g_Y^4$;  
 in \eq(31), $g_s^2 \to g_s^2 / (4\pi)^2$; 
 in \eq(33), ${g_3'}^2 \to {g_3'}^4$ and $2 {h_3'}^2 \to \fr12 {h_3'}^2$; 
 on the 1st line of \eq(35), 
 $
  \ln\frac{\mu^{ }_3}{2 m^{ }_3} \to 
  \ln\frac{\mu^{ }_3}{2 (m_3^2 + \delta m_3^2)^{1/2}}
 $;  
 on the 5th row of 
 \eq(45), $\frac{25}{72} \to \frac{25}{72} \ln$. 

As far as notation goes, we have replaced 
 $g^2 \to g_2^2$, $g'^2 \to g_1^2$,  
 $g_Y^2 \to h_t^2$, $\Lambda\to \bmu$, and $\gamma\to \gammaE$. 
 Because of the trickiness of hypercharge assignments for a general 
 representation, we choose directly Standard Model group theory factors: 
 in the notation of refs.~\cite{gv1,gv2}, 
 \be
  d_\rmii{F} \to 2 \;, \quad
  d_\rmii{A} \to 3 \;, \quad
  C_\rmii{F} \to \fr34 \;, \quad
  C_\rmii{A} \to 2 \;, \quad
  T_\rmii{F} \to \fr12 \;, \quad
  n_\rmii{S} \to 1 \;, \quad
  \Nc \to 3 \;. \la{group}
 \ee
 Only the number of fermion generations, denoted by $\nF$, 
 is left free in our results.
In order to simplify the expressions somewhat, we also substitute
$h^{ }_3 \to g_2^2/4$, $h_3'\to g_1^2/4$.

We disagree with the square brackets
in \eq(198) of ref.~\cite{gv1}. 
In the notation of ref.~\cite{gv1}, 
this term arises from the three-dimensional integral 
\be
 \mathcal{I}^{ }_\rmi{new} \equiv
 \int_{pqr}
 \frac{1}{(p^2 + m_\rmii{D}^2)[(r-p)^2 + m_\rmii{D}^2]
 (q^2 + m_3^2)[(r-q)^2 + m_3^2] r^4}
 \;. 
\ee
Because of an infrared divergence, the integral needs to be carried out 
in the presence of dimensional regularization. The infrared divergence
is powerlike so that, in the end, the limit $\epsilon\to 0$ can be 
taken.  The square brackets in \eq(198) of ref.~\cite{gv1} correspond
to the combination 
$
 [...] \equiv 24(4\pi)^3 m_\rmii{D}^2 m_3^2 \mathcal{I}^{ }_\rmi{new}
$, 
and we believe that this combination has to be replaced as 
 \ba
  & &   \biggl[ \frac{m_\rmii{D}^2}{m^{ }_3} 
    \ln \frac{m^{ }_3 + m^{ }_\rmii{D}}{m^{ }_\rmii{D}} + 
   \frac{m_{3}^2}{m^{ }_\rmii{D}} 
    \ln \frac{m^{ }_3 + m^{ }_\rmii{D}}{m^{ }_{3}}
  - 4 (m^{ }_3 + m^{ }_\rmii{D})  \biggr]
 \nn &  \rightarrow &  
  - \biggl[ \frac{m_\rmii{D}^2}{m^{ }_3} 
    \ln \frac{m^{ }_3 + m^{ }_\rmii{D}}{m^{ }_\rmii{D}} + 
   \frac{m_{3}^2}{m^{ }_\rmii{D}} 
    \ln \frac{m^{ }_3 + m^{ }_\rmii{D}}{m^{ }_{3}}
   + \frac{ m^{ }_3 + m^{ }_\rmii{D}}{2}  \biggr]
   \;.
 \ea
Consequently, a round bracket on the third-but-last line in \eq(35)
of ref.~\cite{gv1} needs to be changed as 
 \ba
  & & 
   g_3^4 C^{ }_\rmii{A} C^{ }_\rmii{F} d^{ }_\rmii{F} \fr13
  \biggl(  
   \frac{m_{3}^2}{m^{ }_\rmii{D}} 
    \ln \frac{m^{ }_\rmii{D} + m^{ }_3 }{m^{ }_{3}} + 
   \frac{m_\rmii{D}^2}{m^{ }_3} 
    \ln \frac{m^{ }_\rmii{D} + m^{ }_3 }{m^{ }_\rmii{D}}  
  \biggr)
   \nn & \rightarrow & 
   g_3^4 C^{ }_\rmii{A} C^{ }_\rmii{F} d^{ }_\rmii{F} \fr16
  \biggl(  
   \frac{m_{3}^2}{m^{ }_\rmii{D}} 
    \ln \frac{m^{ }_\rmii{D} + m^{ }_3 }{m^{ }_{3}} + 
   \frac{m_\rmii{D}^2}{m^{ }_3} 
    \ln \frac{m^{ }_\rmii{D} + m^{ }_3 }{m^{ }_\rmii{D}}  
    +\fr74 \bigl( m^{ }_\rmii{D} + m^{ }_3 \bigr)
  \biggr)
   \;.
 \ea

%%%%%%%%%%%%%%%%%%%%%%%%%%%%%%%%%%%%%%%%%%%%%%%%%%%%%%%%%%%%%%%%%%%%%%%%%%%

\end{document}